\documentclass[14pt]{article}
\usepackage[utf8]{inputenc}
\usepackage{cite}
\usepackage{authblk}
\usepackage{graphicx}
\usepackage{subfig}
\usepackage[margin=1.1in]{geometry}
\usepackage{amsmath}
\usepackage{amssymb}
\usepackage{physics}
\usepackage{cancel}
\usepackage{comment}
\usepackage{tikz}
\usepackage{mwe}
\usepackage{mathtools}
\RequirePackage[colorlinks,citecolor=blue,urlcolor=magenta,linkcolor=blue]{hyperref}
\allowdisplaybreaks
\labelformat{section}{Section #1} 
\labelformat{subsection}{Section #1} 
\labelformat{subsubsection}{Section #1}
\labelformat{subsubsubsection}{Section #1}
\labelformat{equation}{Eq.~(#1)} 
\labelformat{subfigure}{Fig.~\thefigure#1} \labelformat{table}{Table~#1} 
\labelformat{appendix}{Appendix #1}

\usepackage{geometry}
\usepackage{hyperref}
\definecolor{darkslateblue}{rgb}{0.28, 0.24, 0.55}

\makeatletter
\def\@fnsymbol#1{\ensuremath{\ifcase#1\or \ddagger\or \dagger\or
   \mathsection\or \mathparagraph\or \|\or **\or \ddagger\ddagger
   \or \dagger\dagger \else\@ctrerr\fi}}

\title{\bf Singularity avoidance from path integral}

\author[1]{Ribhu Paul,\footnote{ribhupaul.rp@gmail.com}}
\author[1]{Sumanta Chakraborty\footnote{tpsc@iacs.res.in}}
\affil[1]{{\small{{{School of Physical Sciences \authorcr
Indian Association for the Cultivation of Science, Kolkata - 700032, India}}}}}

\date{}

\begin{document}

\maketitle

\begin{abstract}
We have demonstrated that the wavefunction describing the quantum nature of the spacetime inside the black hole horizon, vanishes near the singularity, using the path integral formalism. This is akin to the DeWitt criterion, applied to the interior of a Schwarzschild black hole. To achieve the same we have expressed the interior of a Schwarzschild black hole as a Kantowski-Sachs spacetime and have applied the minisuperspace formalism to determine an exact form of the propagator, and hence the wavefunction near the singularity, using path integral over the reduced phase space. It is to be emphasized that our result is exact and not a saddle point approximation to the path integral.
\end{abstract}
\section{Introduction}

Black holes are the most fascinating, while at the same time one of the most bizarre objects found in our surrounding universe. These are ubiquitous, generically arise after the death of massive stellar structures, and most importantly simplest objects of all. Predominantly, black holes are characterized by three parameters alone, namely the mass $M$, the electric charge $Q$, and the angular momentum $J$ \cite{MTW, Ruffini:1971bza, Bekenstein:1995un, Herdeiro:2015waa, Sotiriou:2015pka}. Despite their simplicity, black holes harbor event horizons, which have rich geometrical structures, and act as one-way membranes, hiding dearly the spacetime beneath the event horizon from the rest of the universe. The horizon itself acts as a thermal system with an entropy proportional to its area and temperature proportional to its surface gravity. The association of thermodynamic properties to black hole horizons\cite{Gibbons:1976ue, Bekenstein:1973ur, Bardeen:1973gs, Hawking:1976de, Wald:1999vt, Padmanabhan:2003gd}, and generically to null surfaces \cite{Chakraborty:2015hna, Chakraborty:2015aja, Chakraborty:2019doh} is considered as the most prominent link to quantum theories of gravity \cite{Bombelli:1986rw, Ashtekar:1997yu, Bousso:2002ju}.   

Amidst all of these intriguing and important findings, there is one persistent issue associated with black hole geometries, namely the existence of a central singularity, which the horizon hides from the outside world. At the singular point the classical notion of spacetime --- which, \textit{per se}, is a smooth, time-orientable, Lorentzian manifold $\mathcal{M}$ equipped with a metric, breaks down. In particular, at the singularity, geodesics end abruptly without any smooth extension, which is physically unsatisfactory \cite{PhysRevD.107.044016, PhysRevLett.14.57}. It is therefore believed that quantum theories of gravity will remove this singular region/point, and make spacetime smooth everywhere \cite{Ashtekar:2005qt, Engelhardt:2016kqb, Brahma:2014gca}. Some progress has been made in this direction from string theory --- where both `resolved' and `excised' singularities exist \cite{Natsuume:2001ba}, in the context of connection dynamics\cite{Brunnemann:2005in, Blanchette_2021}, loop quantum gravity \cite{Ashtekar:2018lag, Cartin:2006yv, Corichi:2015xia, Joe:2014tca}, and several others \cite{Kuntz:2019lzq}. As no complete theory of quantum gravity is in sight, one can still address the above issue using the canonical quantization approach \cite{Hajicek:1992mx, Hajicek:2000mh}. Akin to quantum mechanics and field theory, one writes down a Schr\"{o}dinger-like equation based on the Hamiltonian for the theory of gravity, known as the Wheeler-DeWitt equation \cite{PhysRev.160.1113}. Solution of which provides the wavefunctional of the universe in the superspace of metric and its conjugate momentum. Hence, the problem of singularity can in principle be addressed if the wavefunction of the universe near the singularity vanishes, which is also known as the DeWitt criterion \cite{PhysRev.160.1113}. This would simply mean that singular three-geometries, predicted by classical theories of gravity, are non-existent in the quantum domain. Intriguingly, the vanishing of the wavefunction near the classical singularity has been observed in several works, in the context of minisuperspace approach in quantum theories, e.g., for a classical black hole, as well as for collapsing matter fields forming a black hole \cite{Brahma_2022, Perry:2021mch, Kiefer:2019bxk, Kiefer:2019csi, Kiefer:2019cud}. Implying that quantum gravity indeed resolves black hole singularities. We would like to emphasize that the vanishing of the wavefunction is a necessary condition for singularity avoidance, but is not sufficient (see \cite{Kiefer:2019csi} for details). Furthermore, the DeWitt criteria, at least for the minisuperspace approach, is independent of the probabilistic interpretation of the wavefunction and hence none of the conceptual issues associated with the probabilistic interpretation of the wavefunctional, namely the absence of a well-defined \textit{Lebesgue} measure, or the problem of time, affects the singularity avoidance (for qualitative details, see \cite{article, Anderson_2012}). In this work also, we remain within the paradigm of minisuperspace and hence the DeWitt criteria has a well defined meaning.

There have been several attempts to demonstrate the vanishing of the wavefunctional by solving the Wheeler-DeWitt equation on a Schwarzschild background \cite{Brahma_2022, Bouhmadi-Lopez:2019kkt, Perry:2021mch}. However, the complexity of the equation does not allow for a solution in a general setting, and hence it makes sense to solve the equation in the minisuperspace formalism. Motivated by the facts that the minisuperspace approach had worked extremely well in the context of quantum cosmology, and the spacetime inside the horizon of a Schwarzschild black hole is time-dependent and expressible as Kantowski-Sachs spacetime \cite{Kantowski:1966te}, the wavefunction in the minisuperspace inside the horizon can be determined and it indeed vanishes near the singularity \cite{Kiefer:2019cud, Kiefer:2019bxk, Kiefer:2019csi}. 

Besides solving for the Wheeler-DeWitt equation, there is an alternative approach, namely the path integral formalism (for application to minisuperspace, see \cite{PhysRevD.39.2206, PhysRevD.40.1868, PhysRevD.42.3997, Ailiga:2023wzl}), which also provides the wavefunction in the context of quantum cosmology and thereby allows an exploration of the singular regions in the geometry. The path integral approach is more preferred than solving the Wheeler-DeWitt equation since, the path integral approach depends on the choice of the boundary terms in the gravitational action, which has significance to the gravito-thermodynamic connection and also in the stability of the saddle points in the path integral. Interestingly, there have been several developments in the quantum cosmology sector with the application of the \textit{Picard-Lefschetz} theory in the Lorentzian path integral approach \cite{PhysRevD.95.103508, Lehners_2023, mondal2023lorentzian} and it is expected that some of these results pertaining to the choice of boundary terms in the gravitational action will also be relevant for determining the path integral of Kantowski-Sachs universe, which eventually will apply to the interior of Schwarzschild spacetime. This motivates us to apply the path integral technique to the interior of the Schwarzschild black hole, as it can be described by a cosmological model and thereby the available techniques of Lorentzian quantum cosmology for inflation \cite{Mondal:2022gyp} and bouncing scenario \cite{Rajeev:2021lqk, Rajeev:2021yyl} can be directly applicable. Therefore, we wish to complement the previous approaches in this direction \cite{Brahma_2022,Bouhmadi-Lopez:2019kkt,Kan:2022ism}, which are based on the Wheeler-DeWitt equation, by demonstrating that the vanishing of wavefunction near the singularity can also be arrived at through the path integral approach, with an appropriate choice of the boundary terms in the gravitational action. This would suggest that the classical singularity plaguing the black hole spacetimes can be avoided by invoking the principles of quantum mechanics and this may provide a potential solution to the breakdown of classical physics.

The paper is organized as follows: In \ref{Sch_int} we present the interior solution of the Schwarzschild black hole and relate it to the geometry described by the Kantowski-Sachs universe. Subsequently, we express the gravitational action in the minisuperspace approximation for the Kantowski-Sachs spacetime, which is used in \ref{path_int} to determine the gravitational path integral. The result of the path integral analysis has been used to determine the asymptotic form of the wavefunction at the central singularity in \ref{wavefnl_sing}, which explicitly demonstrates the vanishing of the wavefunction near the singularity. We end with a discussion of our results. Relevant computations have been delegated to the appendix for completeness.

\emph{Notations and Conventions:} We have used the mostly positive signature convention throughout this work, such that the Minkowski metric in the Cartesian coordinates become $\textrm{diag.}(-1,+1,+1,+1)$. We have also set the fundamental constants to unity, i.e., $c=1=G$, but will keep the $\hbar$ throughout our analysis. 

\section{Interior geometry of a Schwarzchild black hole}\label{Sch_int}

In this section, we shall revisit the interior geometry of a Schwarzschild black hole, since our primary concern is the path integral (functional) quantization of the Schwarzschild interior. The starting point is certainly the exterior Schwarzschild geometry, which is defined on a manifold $\mathbb{R}^{2}\times \mathbb{S}^2$, where one of the real coordinates (time) can take all possible values, while the radial coordinate is restricted between $(r_{+},\infty)$, with $r_{+}=2M$. For brevity, we shall stick to the notation $r_{+}$ to denote the location of the Schwarzschild horizon. Thus, the line element describing the exterior geometry of the Schwarzschild spacetime becomes,
\begin{equation}
ds_{\rm ext}^2=-\left(1-\frac{r_{+}}{\mathfrak{r}}\right)d{\mathfrak{t}}^2 
+\left(1-\frac{r_{+}}{\mathfrak{r}}\right)^{-1}d\mathfrak{r}^2 
+\mathfrak{r}^2d\Omega^2_{2}~.
\end{equation}
The interior solution can be obtained by imposing the following heuristic transformation (for a detailed discussion, see \cite{Doran_2007}): $\mathfrak{r} \rightarrow \mathcal{T}$ and $\mathfrak{t} \to r$, where $\mathcal{T} \in [0,r_{+}]$. This results in the following metric, describing the interior solution of the Schwarzschild black hole \cite{Doran_2007},
\begin{equation}
ds_{\rm int}^{2}=-\left(\frac{r_{+}}{\mathcal{T}}-1\right)^{-1}d\mathcal{T}^2 + \left(\frac{r_{+}}{\mathcal{T}}-1\right)d{r}^2+\mathcal{T}^2d\Omega^2_{2}~.
\end{equation}
As evident, the metric is now a function of the `time' coordinate $\mathcal{T}$ and thus can be mapped to the cosmological setting. This is the prime reason for employing path integral techniques to the interior of a Schwarzschild black hole since such techniques have been well-studied and also, they provide sensible answers in the context of cosmology. To reduce the above line element to a more sensible form, we can introduce a new time coordinate $t$, such that,
\begin{equation}
\mathcal{T}({t})=r_{+}\cos ^2\left(\frac{{\pi t}}{2}\right)~,
\end{equation}
and hence the interior metric takes the following form
\begin{equation}\label{interior1}
ds_{\rm int}^2=-\pi^2r_{+}^2 \cos ^4\left(\frac{{\pi t}}{2}\right) d {t}^2+\tan ^2 \left(\frac{{\pi t}}{2}\right) d{r}^2 + r_{+}^2 \cos ^4 \left(\frac{{\pi t}}{2}\right)d\Omega_2^2~.
\end{equation}
The above metric can be framed into the Kantowski-Sachs model, which differs from Bianchi cosmological models in terms of the metric structure, symmetries, and the way anisotropy is manifested and, is particularly useful for describing homogeneous and anisotropic universes. The existence of anisotropy is also evident from \ref{interior1}, as the scale factor in front of the radial coordinate $r$ is different from the scale factor in front of the angular coordinates. To show the anisotropy explicitly and to cast the above metric in a form convenient for the path integral analysis, we make the following identifications,
\begin{equation}
e^{X}\equiv \tan \left(\frac{{\pi t}}{2}\right)~,
\quad
e^{Y}=\frac{1}{2}\sin \left({{\pi t}}\right)~,
\quad
\mathcal{N}({t})=2\pi r_{+}\csc(\pi t)~.
\end{equation}
This leads to the following metric for the interior of the Schwarzschild spacetime, which is a modified version of the Kantowski-Sachs metric \cite{Brahma_2022},
\begin{equation}\label{kantsachs}
ds_{\rm int}^2=-\mathcal{N}^2 e^{4Y-2X}dt^2 + e^{2X} dr^2 + {r_{+}^2 e^{2Y-2X}}d\Omega_2^2~.
\end{equation}
Unlike the exterior geometry, here the spatial section has topology $\mathbb{S}^1 \times \mathbb{S}^2$ which is crucial for a well-defined path-integral\cite{Fanaras:2022twv}. Given the above metric, we notice that the horizon is located at $t\to 0$, which corresponds to $(X\to -\infty, Y\to -\infty)$. On the other hand, the central singularity is located at $t\to 1$, which in terms of the new scale factor yields $(X\to \infty, Y\to -\infty)$. Interestingly, given the sinusoidal functions, the two scale factors $X$ and $Y$ in the classical solution are related by the following constraint: $Y=-\ln(2\cosh X)$. It is important to note that, although the classical scale factors are constrained by this relation, the functional quantization allows for all possible $X$ and $Y$ functions, without any reference to the classical constraint.

Given the above metric, one can now express the gravitational action in terms of the two degrees of freedom $X$ and $Y$, considered here. This corresponds to the mini-superspace approach, where instead of expressing the gravitational action in terms of all the gravitational degrees of freedom, one first imposes symmetry in the background metric itself and reduces the number of degrees of freedom, before expressing the action in terms of these variables. In terms of the variables $X$ and $Y$, the Einstein-Hilbert action functional, along with the Gibbons-Hawking-York boundary term becomes
\begin{equation}\label{gravaction}
\begin{aligned}
\mathcal{A}&=\frac{1}{2\kappa}\int_{\mathcal{M}} d^4x~\sqrt{-g}R+\frac{1}{\kappa}\int_{\partial\mathcal{M}}{d^3x}~\sqrt{|h|}K
\\
&=\frac{\mathcal{V}_R}{\kappa}~\int_{0}^{1}~d{t}~~\left[\frac{\mathcal{N}}{r_{+}^2}e^{2Y}+\frac{1}{\mathcal{N}}(X'^2-Y'^2)\right]~.
\end{aligned}
\end{equation}
Here, $\kappa=8\pi$, $h$ is the determinant of the induced metric on the hypersurface $\partial \mathcal{M}$ and $K$ is the associated extrinsic curvature. On the mini-superspace `prime' denotes the derivative with respect to the time coordinate $t$, which takes value in the range $(0,1)$. Note that $t=0$ corresponds to the horizon, while $t=1$ is the singularity. This shows that a particle crossing the horizon of a Schwarzschild black hole will reach the singularity in a finite time in coordinate $t$. The above action also depicts the usefulness of the above parameterization for the spacetime metric, since the kinetic terms of both of the scale factors in the action have a quadratic nature, which shall be helpful in the path integral approach.

Given the above action, the momenta conjugate to the scale factors $X$ and $Y$ become, $P_{X}=(2X'/\mathcal{N})$, and $P_{Y}=(2Y'/\mathcal{N})$, respectively. The corresponding Hamiltonian is obtained by expressing the action functional in the canonical form as,
\begin{equation}
\mathcal{A}=\frac{\mathcal{V}_R}{\kappa}\int_0^1 dt~\left(P_X X'+P_Y Y'-\mathcal{N}\mathcal{H}\right)~,
\end{equation}
where the Hamiltonian $\mathcal{H}$ takes the following form,
\begin{equation}\label{hamiltonian}
\mathcal{H}=\frac{P_X^2}{4}-\frac{P_Y^2}{4}-\frac{e^{2Y}}{r_{+}^2}~.
\end{equation}
Note that, as far as the momentum terms are considered, the Hamiltonian as well is quadratic, while the potential for the $Y$ scale factor shows the exponential behavior. One can check that owing to the time reparametrization invariance of the gravitational action, the on-shell Hamiltonian weakly vanishes and forms a primary first-class constraint. Moreover, the Hamiltonian also nicely separates into a $X$-dependent and a $Y$-dependent part. This feature is present in the action functional as well, and the above action can also be neatly separated into a $X$ dependent part and a $Y$ dependent part, respectively, such that: $\mathcal{A}[X,Y;\mathcal{N}]=\mathcal{A}^x[X;\mathcal{N}]+\mathcal{A}^y{[Y;\mathcal{N}]}$. Here $\mathcal{A}^{x}$ depends on $X$ alone and $\mathcal{A}^{y}$ depends on $Y$ alone, having the following forms,
\begin{align}\label{eq10}
\mathcal{A}^{x}[X;\mathcal{N}]&=\frac{\mathcal{V}_R}{\kappa}~\int_{0}^1~d{t}\left(\frac{~X'^2}{\mathcal{N}}\right)~;
\qquad
\mathcal{A}^y[Y;\mathcal{N}]=\frac{\mathcal{V}_R}{\kappa}~\int_{0}^1~d{t}\left[\frac{\mathcal{N}}{r_{+}^2}e^{2Y}-\frac{~Y'^2}{\mathcal{N}}\right]~.
\end{align}
The variation of the action functional $\mathcal{A}^{x}$ with respect to the scale factor $X$ yields the free particle equation of motion and requires $X$ to be fixed at the endpoints, while the action functional $\mathcal{A}^{y}$ yields a particle moving in an exponential potential, which can be mapped to the Morse potential, useful for performing path integral. We would like to emphasize that we are using the Gibbons-Hawking-York covariant boundary term for evaluating the path integral. 
\section{Gravitational path integral inside a Schwarzschild black hole}\label{path_int}

The gravitational path integral seeks to capture the quantum properties of the gravitational field by integrating over all spacetime 4-geometries (without any restriction on the bordism class). However, in generic circumstances such a path integral cannot be evaluated explicitly, rather one needs to invoke various symmetry properties to arrive at a symmetry-reduced form of the path integral. Here, we are interested in the quantum properties of the Schwarzschild spacetime near the singularity and for that purpose, we have expressed the interior geometry of the Schwarzschild spacetime in terms of an anisotropic cosmological model. This is because the gravitational path integral in cosmology can be worked out explicitly within the minisuperspace approximation and this enables us to study the quantum properties of the interior of Schwarzschild spacetime. We thus have considered the external space-time geometry to be given by the Schwarzschild spacetime, while the internal geometry is described by three functions---the lapse function $\mathcal{N}$ and the two scale factors $X$ and $Y$. Suppose the wavefunction describing the gravitational field at the horizon is $\Psi_{\rm +}$, while that near the singularity is $\Psi_{\rm s}$. Then, $\Psi_{\rm s}$ can be obtained by evolving the horizon wavefunction $\Psi_{+}$ by the path integral kernel of the metric inside the Schwarzschild horizon. Given the fact that the interior of Schwarzschild spacetime has been framed into a Kantowski-Sachs model, the full infinite-dimensional configuration space of the gravitational field is reduced to a finite-dimensional subspace involving the scale factors and the lapse function, such that the final wavefunction $\Psi_{\rm f}$ inside the Schwarzschild BH can be obtained from the initial wavefunction near the horizon, namely $\Psi(x_{+})$ as
\begin{equation}
\Psi_{\rm f}=\int dx_{+}\int\mathcal{D}g_{\mu \nu}~ e^{{i\mathcal{A}[g]}/\hbar}~\Psi(x_{+}) ~\to ~\int dx_{+}\int \mathcal{D}  X\int \mathcal{D}Y \int \mathcal{D}\mathcal{N}~e^{{i\mathcal{A}[X,Y;\mathcal{N}]}/\hbar}\,\Psi(x_{+})~,
\end{equation}
where, $x_{+}$ denotes collectively a set of relevant dynamical variables determining the wavefunction at the horizon, and the path integral is over the three metric degrees of freedom, the scale factors $X$ and $Y$, as well as the lapse function $\mathcal{N}$.

Evaluation of the above path integral is a bit more involved due to several reasons. Firstly, the diffeomorphism invariance of the gravitational action makes the measure of the path integral ill-defined (arising from over counting of geometric configurations), and it requires choosing a gauge in order to fix the resulting ambiguity. Secondly, the chosen gauge should also satisfy the condition that every possible geometry can be transformed into a configuration that adheres to the gauge. This can be achieved by the following gauge choice
\begin{equation}
{\mathcal{N}'}=G(X,Y,P_X,P_Y,\mathcal{N})~,
\end{equation}
along with the following gauge fixing action, with a Lagrange multiplier $\lambda$,
\begin{equation}
\mathcal{A}_{\rm GF}=\int_{0}^1dt~\lambda({\mathcal{N}'}-G)~,
\end{equation}
to be added to the gravitational action $\widetilde{\mathcal{A}}$. Therefore, the action depends on the gauge fixing function $G$, and to ensure the independence of the path integral on the choice of the gauge fixing function, it is necessary to use the BFV formalism \cite{BATALIN1977309}. In this formalism, one introduces additional anti-commuting ghost fields and their conjugate momenta. The exact form of the action consisting of the ghost fields and their conjugate momentum can be found in \cite{PhysRevD.38.2468}, in the context of minisuperspace cosmology, which we shall not repeat here. 

After all the dust settles down, one chooses the gauge-fixing function, such that $\mathcal{N}'=0$ (BFV gauge choice), which further simplifies the path integral \cite{Lehners_2023}. Therefore, the final wavefunction inside the horizon is given by,
\begin{equation}
\Psi_{\rm f}= \int dx_{+} \int_{\mathbb{R}} {d\mathcal{N}} \int \mathcal{D}X \int \mathcal{D}Y e^{i\mathcal{A}[X,Y;\mathcal{N}]/\hbar}\,\Psi(x_{+})~,
\end{equation} 
where, it has to be mentioned that, the ordinary integration over $\mathcal{N}$ arises from the well-defined Liouville measure of the extended phase space.

Using the decomposition of the action $\mathcal{A}$ into $X$ dependent and $Y$ dependent pieces, the total path integral neatly separates into individual pieces, and the final wavefunction becomes 
\begin{equation}\label{horsingprop}
\Psi_{\rm f}=\int dx_{+} \mathcal{G}(\textrm{f};x_{+})\Psi(x_{+})~,
\quad 
\mathcal{G}(\textrm{f};x_{+})=\frac{1}{r_{+}}\int_{\mathbb{R}} d\mathcal{N}\int \mathcal{D}X e^{i\mathcal{A}^x[X;\mathcal{N}]/\hbar}\int \mathcal{D}Y e^{i\mathcal{A}^y[Y;\mathcal{N}]/\hbar}~.
\end{equation}
Therefore, determining the final wavefunction $\Psi_{\rm f}$ reduces to determining the kernel $\mathcal{G}(\textrm{f};x_{+})$, which involves finding the path integrals over the scale factor $X$ and $Y$, separately, and also integrating the lapse function $\mathcal{N}$ over the real line. The path integral over $X$ can be viewed as the kernel of a free particle, while the path integral over the variable \(Y\) corresponds to a specific form of the Morse potential, as we shall explore. 

\subsection{Path integral over X: free particle kernel with Dirichlet-Dirichlet boundary condition}

The action functional over the scale factor $X$ resembles that of a non-relativistic free particle, which has a well-defined path integral.
As a starting point, we express the scale factor along the $x$ direction as $X(t)=x_{\rm cl}(t)+\eta(t)$, where $x_{\rm cl}$ is the on-shell scale factor subjected to the Dirichlet-Dirichlet boundary condition, with $X(t=1)=x_{1}=x_{\rm cl}(t=1)$, and $X(t=0)=x_{0}=x_{\rm cl}(t=0)$. 
Thus, we obtain the following boundary conditions for the `quantum' position variable $\eta(t)$, 
\begin{equation}\label{eq27}
{\eta}(t=0)=0;\qquad  \eta(t=1)=0\,.
\end{equation}
Given the above decomposition of the scale factor $X$, we can express the corresponding action $\mathcal{A}^{x}$ in the following manner, 
\begin{align}
\mathcal{A}^{x}&=\frac{\mathcal{V}_R}{\kappa}\left[\int_{0}^1~d{t}~\left(\frac{x_{\rm cl}'^2+\eta'^{2}+2x_{\rm cl}'\eta'}{\mathcal{N}}\right)\right]
\nonumber
\\
&=\frac{\mathcal{V}_R}{\kappa}\left[\int_{0}^1~d{t}~\left(\frac{x_{\rm cl}'^2}{\mathcal{N}}\right)\right]+\frac{\mathcal{V}_R}{\kappa}\int_{0}^1~d{t}\left(\frac{\eta'^{2}}{\mathcal{N}}\right)~,
\end{align}
where, we have used the boundary conditions on $\eta$, as elaborated in \ref{eq27}, as well as the fact that $x_{\rm cl}''=0$. The first part of the action is dependent on the classical scale factor $x_{\rm cl}$ alone and is denoted by $\mathcal{A}^{x}_{\rm cl}$, while the second part is a quadratic action for $\eta$, referred to as $\mathcal{A}_{\eta}^{x}$. Hence, the path integral over $X$ can now be carried out as follows where we introduce $\mathcal{G}^x(x_1,x_0;\mathcal{N})$ as
\begin{equation}\label{xint}
\mathcal{G}^x(x_1,x_0;\mathcal{N})\equiv\int_{X(t=0)=x_0}^{X(t=1)=x_1} \mathcal{D}X ~e^{i\mathcal{A}^x/\hbar}
=e^{i\mathcal{A}^{x}_{\rm cl}/\hbar}~\int_{\eta(t=0)=0}^{{\eta}(t=1)=0}\mathcal{D}{\eta} ~e^{i\mathcal{A}^{x}_{\eta}/\hbar}
=\sqrt{\frac{1}{\pi i \hbar_{\rm eff}\mathcal{N}}}~e^{i\mathcal{A}^x_{\rm cl}/\hbar}~
\end{equation}
where, we have introduced the `effective' Planck's constant $\hbar_{\rm \rm{eff}}={\kappa \hbar}/{\mathcal{V}_R}$. It is to be noted that, the factor $(1/\sqrt{\mathcal{N}})$ arises in the path integral due to the imposition of the Dirichlet-Dirichlet boundary condition, which appears naturally from the covariant Gibbons-Hawking-York boundary term.  
We now need to determine the classical action, for which we need the classical solution that adheres to the boundary criteria $x_{\rm cl}(t=0)=x_{0}$, and $x_{\rm cl}(t=1)=x_{1}$---which turns out to be
\begin{equation}
x_{\rm cl}(t)=\left(x_{1}-x_{0}\right)t+x_{0}~.
\end{equation}
Substituting the above solution in the classical/on-shell action $\mathcal{A}^{x}_{\rm cl}$ gives,
\begin{align}
\mathcal{A}^{x}_{\rm cl}&=\frac{\mathcal{V}_R}{\kappa}\int_0^1~dt~ \left(\frac{x_{\rm cl}'^2}{\mathcal{N}}\right)
=\frac{\mathcal{V}_R}{\kappa}\left(\frac{(x_{1}-x_{0})^{2}}{\mathcal{N}}\right)~.
\end{align}
Now, that all the necessary components are available, the path integral over the scale factor $X$ leads to the following expression,
\begin{equation}
\mathcal{G}^x(x_1,x_0;\mathcal{N})=\left(\sqrt{\frac{1}{\pi i \hbar_{\rm eff}\mathcal{N}}}\right)~\exp\left[\frac{i(x_{1}-x_{0})^{2}}{\mathcal{N}\hbar_{\rm eff}}\right]~.
\end{equation}
This is the expression we are after, yielding path integral of the scale factor $X$. 
Keeping our later purpose in mind, it is instructive to express the above path integral kernel for the scale factor $X$ as, 
\begin{equation}
\mathcal{G}^x(x_1,x_0;\mathcal{N})=\int dp_{0}e^{-ip_{0}x_{0}/\hbar_{\rm eff}}
\left[\frac{1}{2\pi \hbar_{\rm eff}}e^{ip_{0}x_{1}/\hbar_{\rm eff}}e^{-i\mathcal{N}p_{0}^{2}/4\hbar_{\rm eff}}\right]~.
\end{equation}
The usefulness of the above expression can be understood from the fact that the Lapse function $\mathcal{N}$ appears in the exponent, rather than as an overall factor involving $(1/\sqrt{\mathcal{N}})$. This fact will be helpful while determining the wavefunction describing inside of the Schwarzschild BH. 
\subsection{Path integral over $Y$: kernel for the Morse-like potential}

The path integral over $Y$ is non-trivial due to the appearance of the exponential potential. Such potential appears in problems related to molecular vibration and is known as the Morse potential. The Green's function corresponding to one-dimensional Morse potential has previously been obtained in the context of functional integration in \cite{PhysRevD.28.2689}, which we closely follow here.

For evaluating the path integral we shall consider the Dirichlet boundary condition on both the boundaries of $Y$ such that,
\begin{equation}
Y(t=0)=y_0~;\qquad Y(t=1)=y_1~.
\end{equation}
It is advantageous to express the above path integral in the phase space, where it takes the following form
\begin{equation}
\mathcal{G}^{y}(y_{1},y_{0};\mathcal{N})\equiv \int \mathcal{D}Y~ e^{i{\mathcal{A}}^y/\hbar}=\int \frac{\mathcal{D}Y \mathcal{D}P_Y}{2\pi \hbar_{\rm eff}}~\exp\left[\frac{i}{\hbar_{\rm \rm{eff}}}\int_{0}^{1}dt~\left\{P_Y Y'+\frac{\mathcal{N}}{4}P_Y^2+\frac{ \mathcal{N}}{r_{+}^2}e^{2Y}\right\}\right]~,
\end{equation}
where we have introduced the momentum $P_{Y}$, conjugate to the scale factor $Y$ as, 
\begin{equation}
P_{Y}=-\frac{2}{\mathcal{N}}Y'~.
\end{equation}
Computing the path integral requires performing a point transformation to a new set of canonical chart---($\varsigma, P_{\varsigma}$), related to the old canonical chart $(Y, P_{Y})$ through the following relation
\begin{equation}
Y=2\ln \varsigma~;\qquad P_{Y}=\frac{1}{2}\varsigma P_{\varsigma}~,
\end{equation}
generated by the following generating function of the second kind: $F_2(Y, P_{\varsigma})=e^{Y/2}~P_{\varsigma}$. In terms of these new phase space variables, the above path integral takes the following form,
\begin{equation}
\mathcal{G}^{y}(y_{1},y_{0};\mathcal{N})=\frac{1}{2}\varsigma_1 \int \frac{\mathcal{D}\varsigma \mathcal{D}P_{\varsigma}}{2\pi \hbar_{\rm eff}}~\exp\left[\frac{i}{\hbar_{\rm \rm{eff}}}\int_0^1 dt~\left\{P_{\varsigma}\varsigma'+\frac{\mathcal{N}}{16}\varsigma^2 P_{\varsigma}^2+\frac{ \mathcal{N}}{r_{+}^2} \varsigma^4\right\}\right]~,
\end{equation}
where the factor $\frac{1}{2}\varsigma_1$ comes from the Jacobian of the above phase space transformation. Subsequently, performing a simple re-scaling of $t$ to $\mathcal{N}t$ reduces the above path integral to the following form,
\begin{equation}
\mathcal{G}^{y}(y_{1},y_{0};\mathcal{N})=\frac{1}{2}\varsigma_1 \int \frac{\mathcal{D}\varsigma \mathcal{D}P_{\varsigma}}{2\pi\hbar_{\rm eff}}~\exp\left[\frac{i}{\hbar_{\rm \rm{eff}}}\int_0^\mathcal{N} dt~\left\{P_{\varsigma}\varsigma'+\frac{1}{16}\varsigma^2 P_{\varsigma}^2+\frac{1}{r_{+}^2} \varsigma^4\right\}\right]~.
\end{equation}
Following \cite{PhysRevD.28.2689}, we introduce an auxiliary time variable $s$, monotonic in $t$, known as the Duru-Kleinert reparametrization, such that $dt=ds/\varsigma(s)^2$, yielding
\begin{equation}
t=\int^s \frac{d\mathfrak{s}}{\varsigma^2(\mathfrak{s})}~.
\end{equation}
Note that the initial and final points, namely $t=0$ and $t=\mathcal{N}$, are denoted by $s=s_{0}$ and $s=s_{1}$, respectively. Defining, $S\equiv s_{1}-s_{0}$, it follows that the lapse function can be expressed as, 
\begin{equation}
\mathcal{N}=\int^S \frac{ds}{\varsigma(s)^{2}}~.
\end{equation}
Transforming the newly defined time coordinate $s$, and imposing the above constraint between the lapse function and the new time coordinate $s$, the path integral over the scale factor $Y$ reduces to, 
\begin{equation}
\begin{aligned}
\mathcal{G}^{y}(y_{1},y_{0};\mathcal{N})=&-\frac{1}{2}\varsigma_1\left(\frac{1}{\varsigma_1^2}\right)\int_0^{\infty}~dS~ \delta\left(\mathcal{N}-\int^S_{0}\frac{ds}{\varsigma(s)^{2}}\right)\int \frac{\mathcal{D}\varsigma \mathcal{D}P_{\varsigma}}{2\pi\hbar_{\rm eff}}
\\
&\qquad \qquad \times \exp\left[\frac{i}{\hbar_{\rm \rm{eff}}}\int_0^S ds~\left\{P_{\varsigma}\varsigma'+\frac{1}{16}P^2_{\varsigma}+\frac{1}{r_{+}^2} \varsigma^2\right\}\right]~.
\end{aligned}
\end{equation}
Here, the term $(1/\varsigma^2)$ comes from the normalization of the $\delta$ function and the prime now denotes derivative with respect to the new time coordinate $s$. By applying the delta function constraint, the lapse function $\mathcal{N}$ is \rm{eff}ectively set to a specific value, while the other variable $S$ is allowed to vary independently without being influenced by the constraint imposed by the delta function. This is a crucial feature that allows one to perform the $\mathcal{N}$ integration before the functional integrations. Using the Fourier decomposition of the delta function, we can neatly separate the path integral into a term depending on the lapse function and then an integration over $S$, such that
\begin{equation}
\mathcal{G}^{y}(y_{1},y_{0};\mathcal{N})=-\frac{1}{2\varsigma_1}\int_{-\infty}^{\infty} \frac{d\lambda}{2\pi} e^{i\lambda \mathcal{N}}\int_0^{\infty}dS~\int \frac{\mathcal{D}\varsigma \mathcal{D}P_{\varsigma}}{2\pi\hbar_{\rm eff}}\exp\left[\frac{i}{\hbar_{\rm \rm{eff}}}\int_0^S ds~\left\{P_{\varsigma}\varsigma'+\frac{1}{16}P^2_{\varsigma}+\frac{1}{r_{+}^2} \varsigma^2-\frac{\lambda \hbar_{\rm \rm{eff}}}{\varsigma^2}\right\}\right]~.
\end{equation}
Integrating over the conjugate momenta $P_{\varsigma}$, the above expression yields,
\begin{equation}
\mathcal{G}^{y}(y_{1},y_{0};\mathcal{N})=-\frac{1}{2}\int_{-\infty}^{\infty}\frac{d\lambda}{2\pi} e^{i\lambda \mathcal{N}}\int_0^{\infty}dS~\int \left(\frac{1}{\varsigma_1}\right){\mathcal{D}\varsigma}\exp\left[\frac{i}{\hbar_{\rm \rm{eff}}}\int_0^S ds\left\{-4{\varsigma'}^{2}+\frac{1}{r_{+}^2}\varsigma^2-\frac{\lambda \hbar_{\rm \rm{eff}}}{\varsigma^2}\right\}\right]~.
\end{equation}
The above path integral for the scale factor $Y$, with a suitable change of variables, can be reduced to the one, that has already been computed in \cite{PhysRevD.28.2689}, and reads,
\begin{equation}\label{prop_mom}
\begin{aligned}
\mathcal{G}^{y}(y_{1},y_{0};\mathcal{N})&=-r_{+}\int_{-\infty}^{\infty}\frac{d\lambda}{2\pi} e^{i\lambda \mathcal{N}}
\int_0^{\infty}d\sigma \left(\frac{4i}{r_{+}\hbar_{\rm eff}\sin{\sigma}}\right)
I_{\frac{2ip_{0}}{\hbar_{\rm \rm{eff}}}}\left(\frac{4ie^{\frac{y_1+y_0}{2}}}{r_{+}\hbar_{\rm eff} \sin{\sigma}}\right)\exp\left[-\frac{2i}{r_{+}\hbar_{\rm eff}}\left(e^{y_1}+e^{y_0}\right)\cot \sigma\right]~,
\end{aligned}
\end{equation}
where, we have defined, $\sigma\equiv (S/2r_{+})$. This is the final form of the path-integral over the $y$ direction, which we will use next to determine the wavefunction near the Schwarzschild singularity.

\subsection{Determining wavefunction inside a Schwarzchild black hole}

In order to determine the wavefunction inside a Schwarzschild black hole, we can use \ref{horsingprop}, along with the path integral over the scalar factors $X$ and $Y$ derived in the previous sections. For that purpose, we note that the function $\mathcal{G}(\rm f;x_+)$, defined in \ref{horsingprop}, can be expressed as,
\begin{equation}\label{greeninter}
\mathcal{G}(x_1, y_{1}; x_0, y_0)=\frac{1}{r_{+}}\int_{-\infty}^{\infty}d\mathcal{N}~\mathcal{G}^{y}(y_{1},y_{0};\mathcal{N})\mathcal{G}^x(x_1,x_0;\mathcal{N})~,
\end{equation}
and the final wavefunction becomes an integral over $x_{0}$ and $y_{0}$ of the combination $\mathcal{G}(x_1, y_1; x_0, y_0)\Psi(x_{+})$. Therefore, the final wavefunction depends on the final values of the scale factors $x_{1}$ and $y_{1}$, respectively, and requires knowledge of the function $\mathcal{G}(x_1, y_1; x_0, y_0)$ as well as of the initial wavefunction. We first derive the quantity $\mathcal{G}(x_1, y_1; x_0, y_0)$, and then discuss the choice of the initial wavefunction. 

Substituting the path integral over the scale factor $X$, as well as the path integral over $Y$, both satisfying the Dirichlet-Dirichlet boundary condition, derived in the previous sections, in \ref{greeninter}, we obtain, 
\begin{equation}\label{prop_momentum}
\begin{aligned}
\mathcal{G}(&x_{1},y_{1};x_{0},y_{0})=\frac{1}{r_{+}}\int_{-\infty}^{\infty} d\mathcal{N}
\int_{x_{0}}^{x_{1}} \mathcal{D}X e^{i\mathcal{A}^x[X;\mathcal{N}]/\hbar}
\int_{y_{0}}^{y_{1}}\mathcal{D}Y e^{i\mathcal{A}^y[Y;\mathcal{N}]/\hbar}
\\
&=-\frac{1}{2\pi \hbar_{\rm eff}}\int dp_{0}e^{ip_{0}(x_{1}-x_{0})/\hbar_{\rm eff}}
\int_{-\infty}^{\infty} d\mathcal{N}~e^{-{i\mathcal{N}p_{0}^{2}}/{{4\hbar_{\rm \rm{eff}}}}}\int_{-\infty}^{\infty}\frac{d\lambda}{2\pi} e^{i\lambda \mathcal{N}}
\\
&\qquad \times\int_0^{\infty}d\sigma \left(\frac{4i}{r_{+}\hbar_{\rm \rm{eff}} \sin{\sigma}}\right)
I_{\frac{2ip_{0}}{\hbar_{\rm \rm{eff}}}}\left(\frac{4ie^{\frac{y_1+y_0}{2}}}{r_{+}\hbar_{\rm \rm{eff}} \sin{\sigma}}\right)\exp\left[-\frac{2i}{r_{+}\hbar_{\rm \rm{eff}}}\left(e^{y_1}+e^{y_0}\right)\cot \sigma\right]
\\
&=-\frac{1}{2\pi}\int d\nu e^{i\nu(x_{1}-x_{0})}
\int_0^{\infty}d\sigma~\left(\frac{4i}{r_{+}\hbar_{\rm \rm{eff}} \sin{\sigma}}\right)I_{2i\nu}\left(\frac{4ie^{\frac{y_1+y_0}{2}}}{r_{+}\hbar_{\rm \rm{eff}} \sin{\sigma}}\right)\exp\left[-\frac{2i}{r_{+}\hbar_{\rm \rm{eff}}}\left(e^{y_1}+e^{y_0}\right)\cot \sigma\right]~.
\end{aligned}
\end{equation}
In arriving at the last line, we have first integrated out the lapse function $\mathcal{N}$, yielding a delta function over $\lambda$, and a subsequent integration over $\lambda$ yielding the final form for $\mathcal{G}(x_{1},y_{1};x_{0},y_{0})$, as presented above\footnote{Note that we have performed the integration over the lapse function $\mathcal{N}$ before integrating over $\lambda$. In the present context this does not pose any problem, as we are performing the lapse integral after all the phase space integration have been performed \cite{Banihashemi:2024aal}, and the path integral is being evaluated exactly. If our interest was to approximate the path integral by its value at the saddle points, then the lapse integration needs to be performed at the end. Moreover, since both the lapse integration as well as the $\lambda$ integration are ordinary integration, they can be interchanged. To see this explicitly note that if we perform the integration over $\lambda$ first, yielding $\delta(\mathcal{N})$, the final integration over the lapse function $\mathcal{N}$ would yield identity, and our result will not change.}. We have also defined $\nu\equiv (p_0/\hbar_{\rm \rm{eff}})$. The above integrand has poles at $\sigma=m\pi$, where $m$ is an integer, and since the integration is over the real line, one needs to deform the contour appropriately. For this purpose one makes a Wick-like rotation to the $\sigma$ coordinate by defining $w=-i\sigma$, following \cite{Grosche1,Grosche:1988um}, which amounts to rotate the contour of integration, and then one modifies the variable of integration to shift the locations of the poles from their former position along the contour. This facilitates the integration and as elaborated in \ref{F}, the above integral can be evaluated to the following form,
\begin{equation}\label{prop_theta}
\mathcal{G}=\mathcal{G}^+\Theta(y_0-y_1)+\mathcal{G}^-\Theta(y_1-y_0)~,
\end{equation}
where, $\mathcal{G}^{+}$, associated with the scale factor along the $y$-direction satisfying $y_0>y_1$ (as is the case here, as scale factor $y_{0}$ is related to the horizon and the scale factor $y_{1}$ is inside the horizon, hence necessarily smaller than $y_{0}$) is given by,
\begin{equation}\label{eq49}
{\mathcal{G}}^{+}(x_1,y_1;x_0,y_0)=\int d\nu \frac{2^{2(1+i\nu)}}{2i\sqrt{\pi}r_{+}\hbar_{\rm eff}}e^{{i\nu (x_1-x_{0})}}~\frac{\Gamma\left(\frac{1}{2}+i\nu\right)\Gamma(1+i\nu)}{\Gamma(1+2i\nu)} I_{i\nu}\left(\frac{2e^{y_1}}{r_{+}\hbar_{\rm \rm{eff}}}\right) K_{i\nu}\left(\frac{2e^{y_0}}{r_{+}\hbar_{\rm \rm{eff}}}\right)~.
\end{equation}
The other quantity, namely $\mathcal{G}^{-}$, is obtained by replacing $y_{0}$ and $y_{1}$, respectively, in the expression for $\mathcal{G}^{+}$. Given the above expressions, it follows that the part of $\mathcal{G}^{\pm}$, dependent on the final scale factors $x_{1}$ and $y_{1}$, reads, 
\begin{equation}
\mathcal{G}^+\propto e^{{i\nu x_1}} I_{i\nu}\left(\frac{2e^{y_1}}{r_{+}\hbar_{\rm \rm{eff}}}\right)~;
\qquad 
\mathcal{G}^-\propto e^{{i\nu x_1}} K_{i\nu}\left(\frac{2e^{y_1}}{r_{+}\hbar_{\rm \rm{eff}}}\right)~,
\end{equation}
and hence one can verify that both of them satisfy the Wheeler-DeWitt equation, i.e.,
\begin{equation}
\left[\hbar_{\rm \rm{eff}}^2\left(\frac{\partial^2}{\partial x_1^2}-\frac{\partial^2}{\partial y_1^2}\right)+\frac{4e^{2Y}}{r_+^2}\right]\mathcal{G}^{\pm}(x_1,y_1;x_0,y_0)=0~.
\end{equation}
Here, we have ignored an overall normalization factor, which we will comment on in the next section. This result is exact and follows from the path integral of the Einstein-Hilbert action with appropriate boundary conditions within the mini-superspace formalism applied to Kantowski-Sachs spacetime, which describes the interior of Schwarzschild geometry. 
The validity of this wavefunction can thus be reinforced by the fact that it satisfies the Wheeler-DeWitt equation, as $e^{i\nu x_1}$ and the modified Bessel function are known to be solutions of the Wheeler-DeWitt equations in the context of Kantowski-Sachs universe \cite{Bouhmadi-Lopez:2019kkt}. Given the wavefunction inside the horizon, we now focus on determining the same near the singularity and show that it indeed vanishes in the limit $\mathfrak{r}\to 0$, where $\mathfrak{r}$ is the Schwarzschild radial coordinate.

\section{Wavefunction near the central singularity}\label{wavefnl_sing}

In this section, we will provide an explicit expression for the wavefunction of the Kantowski-Sachs universe, describing the interior of the Schwarzschild black hole, in the vicinity of the Schwarzschild singularity. We have previously presented the final form of the wavefunction as an integral over the kernel $\mathcal{G}(x_1,y_1;x_0,y_0)$. Therefore, the final wavefunction describing the interior of a Schwarzschild black hole can be expressed as, 
\begin{equation}\label{final_2}
\Psi_{\rm f}(y_1,x_1)=\int_{\mathbb{R}} dy_0 \int_{\mathbb{R}}{dx_0}~\mathcal{G}(y_1,x_1;y_0,x_0) \Psi_{\rm i}(y_0,x_0)~,
\end{equation}
which may be rewritten in the following manner:
\begin{equation}
\begin{aligned}
&\Psi_{\rm f}(y_1,x_1)=\int_{\mathbb{R}} dy_0 \int_{\mathbb{R}}{dx_0}~\mathcal{G}^+(y_1,x_1;y_0,x_0)\Theta(y_0-y_1) \Psi_{\rm i}(y_0,x_0)
\nonumber
\\
&\qquad +\int_{\mathbb{R}} dy_0 \int_{\mathbb{R}}{dx_0}~\mathcal{G}^-(y_1,x_1;y_0,x_0)\Theta(y_1-y_0) \Psi_{\rm i}(y_0,x_0)~,
\end{aligned}
\end{equation}
where, $\mathcal{G}^{\pm}(y_1,x_1;y_0,x_0)$ are obtained from \ref{eq49}, and hence the complete determination of the final wavefunction, requires input about the initial wavefunction $\Psi_{\rm i}(y_0,x_0)$.

Note that in the context of the Kantowski-Sachs universe, the location of the singularity translates to the following limit: $Y\to -\infty,~ X\to \infty$. Therefore, only the term $\mathcal{G}^{+}$ in \ref{prop_theta} contributes if we are interested in determining the wavefunction near the singularity (by letting $y_1\to-\infty$ first). Plugging the explicit expression for $\mathcal{G}^{+}$ from \ref{eq49}, after some straightforward algebra, we obtain the following behavior of the wavefunction near the singularity,
\begin{align}\label{eq60}
\Psi_{\rm s}&=A\lim_{\substack{y_1\to -\infty \\ x_1\to \infty}}\int_{\mathbb{R}} dy_0 \int_{\mathbb{R}}{d\nu}\,\frac{2^{2(1+i\nu)}}{2\pi i\sqrt{\pi}}\frac{\Gamma\left(\frac{1}{2}+i\nu\right)\Gamma(1+i\nu)}{\Gamma(1+2i\nu)}e^{ix_{1}\nu}~I_{i\nu}\left(\frac{2}{r_{+}}e^{y_1}\right)~K_{i\nu}\left(\frac{2}{r_{+}}e^{y_{0}}\right)\Psi_{i}(y_{0},\nu)
\nonumber
\\
&\simeq A \lim_{\substack{y_1\to -\infty \\ x_1\to \infty}}\int_{\mathbb{R}} dy_0 \int_{\mathbb{R}}{d\nu}\,e^{ix_{1}\nu}e^{iy_{1}\nu}~\left[r_{+}^{-2i\nu}e^{i\nu y_{0}}\frac{\Gamma(-i\nu)}{i\pi\Gamma(1+i\nu)}+e^{-i\nu y_{0}}\frac{\Gamma(i\nu)}{i\pi\Gamma(1+i\nu)}\right]\Psi_{i}(y_{0},\nu)~,
\end{align}
with, $A=(\pi/r_+\hbar_{\rm eff})$, and $\Psi_{i}(y_{0},\nu)=\int dx_{0}e^{-i\nu x_{0}}\Psi_{i}(y_{0},x_{0})$. Here, we have used the results that for small argument i.e. $z\to 0$, the modified Bessel function can be expressed as $I_{\alpha}(z)\approx \{1/\Gamma(1+\alpha)\}(z/2)^{\alpha}$, and the fact that one can express the other modified Bessel function as: $K_{i\alpha}(az)\approx\frac{1}{2}[(a/2)^{i\alpha}z^{i\alpha}\Gamma(-i\alpha)+(a/2)^{-i\alpha}z^{-i\alpha}\Gamma(i\alpha)]$. We now provide the choice for the initial state, which causes the wavefunction near the singularity to vanish. 

Considering the initial state on the horizon $\Psi_{\rm i}(y_0,\nu)$, to have a spread in the momentum $p_{0}=\nu\hbar_{\rm eff}$, which is conjugate to the scale factor $X$, and assuming the wavefunction to be peaked at a given value of the initial scale factor $y_{0}=\chi$, we have, $\Psi_{i}(y_{0},\nu)=\Phi(\nu)\delta(y_{0}-\chi)$. Given this choice for the initial horizon wavefunction $\Psi_{\rm i}(y_{0},\nu)$, the final wavefunction near the singularity can be obtained from \ref{eq60}, as,
\begin{align}\label{eq601}
\Psi_{\rm s}&\simeq A\lim_{\substack{y_1\to -\infty \\ x_1\to \infty}}\int_{\mathbb{R}}{d\nu}\,
e^{iy_{1}\nu}~\left[r_{+}^{-2i\nu}e^{i\nu (\chi+x_{1})}\frac{\Gamma(-i\nu)}{i\pi\Gamma(1+i\nu)}+e^{-i\nu (\chi-x_{1})}\frac{\Gamma(i\nu)}{i\pi\Gamma(1+i\nu)}\right]\Phi(\nu)~.
\end{align}
Note that the absolute value of the integrand, except the $e^{i\nu y_{1}}$ term, which involves the quantity inside the square bracket multiplied by $\Phi(\nu)$ is always smaller than $|\Phi(\nu)/\nu|$ (using triangle inequality: $|z_{1}+z_{2}|\leq |z_{1}|+|z_{2}|$, for $z_1, z_2 \in \mathbb{C}$). Now, it can be shown that a judicial choice of the initial minisuperspace wavefunction naturally aligns with the DeWitt criteria. For instance, we assume that $\Phi(\nu)$ is a double-Gaussian function defined as $\Phi(\nu) = B\nu^{N}\exp(-\nu^2)$, where $N$ is a positive integer and $B$ is a pure phase function. This function is symmetric/anti-symmetric about $\nu = 0$ depending on the value of $N$, and has maxima (and minima) at $\nu = \pm\sqrt{N/2}$, which can be interpreted as representing the initial wavefunction of the universe, involving both ingoing and outgoing momenta values, as discussed in \cite{Bouhmadi-Lopez:2019kkt}. Under this assumption, the integral of $|\Phi(\nu)/\nu|$ is $L^1(\mathbb{R})$ finite. Thus, using Riemann-Lebesgue lemma\footnote{The Riemann-Lebesgue lemma applies for any $L^{1}$ function on the real line. In particular, if $f\in L^1(\mathbb{R})$, which implies $\int_{\mathbb{R}}dx~|f(x)|<\infty$, then it follows that $\int_{\mathbb{R}}dx~f(x) \exp(-ipx)\to 0$, as $|p|\to\infty$. The fact that the integral vanishes is crucial in depicting vanishing wavefunction near the singularity.}, in the limit $y_1\to-\infty$, it follows that $\Psi_{\rm s}$ vanishes. In the simplest case, i.e. for $N=1$, the choice of $\Phi(\nu)$ can be approximated as the sum of two Gaussian functions, anti-symmetric about the origin.

Therefore, through the expansion of the modified Bessel functions and use of the properties of Gamma functions, along with the Riemann-Lebesgue lemma, it is clear that in the limit $y_{1}\to -\infty$, the wavefunction identically vanishes. This is valid provided the initial wavefunction describes both ingoing and outgoing momentum states. Thus the DeWitt criteria is satisfied at the Schwarzschild singularity, and the wavefunction arising from path integral, describing the interior of a Schwarzschild black hole, goes to zero near the Schwarzschild singularity.  

\section{Discussion}

We have systematically quantized the interior of a Schwarzschild black hole. Since space and time interchange their role inside the black hole horizon, the metric becomes time-dependent and can be expressed as a cosmological spacetime, which in this particular context corresponds to the Kantowski-Sachs spacetime. Assuming that the symmetries of cosmological spacetimes, such as Kantowski-Sachs, hold good in the quantum description of geometry as well, the quantization of the Schwarzschild interior can be described within the framework of minisuperspace cosmology. Since the Kantowski-Sachs model describes anisotropic cosmology, the quantization requires path integral over the scale factors $X$ and $Y$, along the anisotropic directions, along with integration over the lapse function. The path integral over the scale factor $X$ is identical to that of a free particle, while the path integral over the scale factor $Y$ is associated with an exponential potential, which can be converted to a well-known problem in quantum mechanics involving Morse potential. Motivated by the recent results in the path integral description of minisuperspace cosmology, to avoid the branch cut in the integral over the lapse function, we have considered the Dirichlet-Dirichlet boundary conditions over the scale factor $X$, along with Dirichlet-Dirichlet boundary condition over the scale factor $Y$. Importantly, our exploration revealed that the wavefunction of the interior of a Schwarzschild black hole vanishes near the singularity and hence naturally aligns with the DeWitt criteria, for a reasonable choice of the initial minisuperspace wavefunction containing incoming and outgoing momenta. It must be emphasized that in our approach, we have considered a full quantum wavefunction rather than a semi-classical one. This is because we have used the exact path integral wavefunction and not the saddle point approximation to the same. Thus, our result suggests that even in full quantum gravity, the wavefunction of the Schwarzschild geometry vanishes near the singularity, implying its avoidance in the quantum regime, provided the quantum geometry has the same set of symmetries as the classical one.

\section*{Acknowledgements} 
R.P. acknowledges {Vikramaditya Mondal} for valuable discussions and insightful comments. R.P. is funded by the University Grants Commission (UGC), Government of India (Ref: 211610000414). S.C. thanks the Albert Einstein Institute for its warm hospitality, where a part of this work was performed. The visit to the Albert Einstein Institute is funded by the Max-Planck Society through its Max-Planck-India mobility grant. Research of S.C. is supported by the Mathematical Research Impact Centric Support (MATRICS) and the Core research grants from ANRF, SERB, Government of India (Reg. Nos. MTR/2023/000049 and CRG/2023/000934).
\appendix
\labelformat{section}{Appendix #1} 
\labelformat{subsection}{Appendix #1}
\section{Integration involving Bessel function}\label{F}

The integral in \ref{prop_momentum} can be evaluated exactly in terms of Whittaker functions by performing a change of variable --- $w=-i\sigma$ followed by a second transformation $\sinh v=1/\sinh w$ as \cite{Grosche1, Grosche:1988um} 
\begin{equation}\label{eq84}
\begin{aligned}
    &\left(\frac{4i}{r_{+}\hbar_{\rm eff}}\right)\int_0^{\infty}dv~\exp\left[-\frac{2}{r_{+}\hbar_{\rm eff}}(e^{y_1}+e^{y_0})\cosh v\right] I_{2i\nu}\left(\frac{4e^{\frac{y_1+y_0}{2}}}{r_{+}\hbar_{\rm eff}}\sinh v\right)
    \\
    &=\left(\frac{4i}{r_{+}\hbar_{\rm eff}}\right)\frac{r_{+}\hbar_{\rm eff}\Gamma\left(\frac{1}{2}+i\nu\right)}{4e^{\frac{y_1+y_0}{2}}\Gamma(1+2i\nu)}\times W_{0,i\nu}\left(\frac{4e^{y_0}}{r_{+}\hbar_{\rm eff}}\right)M_{0,i\nu}\left(\frac{4e^{y_1}}{r_{+}\hbar_{\rm eff}}\right)
    \\
    &={e^{-\frac{y_1+y_0}{2}}}\frac{i\Gamma\left(\frac{1}{2}+i\nu\right)}{~\Gamma(1+2i\nu)}\times \sqrt{\frac{4}{\pi r_{+}\hbar_{\rm eff}}e^{y_0}} ~K_{i\nu}\left(\frac{2e^{y_0}}{r_{+}\hbar_{\rm eff}}\right) \times 2^{2i\nu +\frac{1}{2}}~\Gamma(1+i\nu)\sqrt{\frac{2}{r_{+}\hbar_{\rm eff}}e^{y_1}}~I_{i\nu}\left(\frac{2e^{y_1}}{r_{+}\hbar_{\rm eff}}\right)
    \\
    &=\frac{4i}{\sqrt{\pi}r_{+}\hbar_{\rm eff}}~2^{2i\nu}~\frac{\Gamma\left(\frac{1}{2}+i\nu\right)\Gamma(1+i\nu)}{\Gamma(1+2i\nu)}\times I_{i\nu}\left(\frac{2e^{y_1}}{r_{+}\hbar_{\rm eff}}\right) \times K_{i\nu}\left(\frac{2e^{y_0}}{r_{+}\hbar_{\rm eff}}\right)~,
\end{aligned}
\end{equation}
when $y_0>y_1$, and we have defined $\nu\equiv (p_0/\hbar_{\rm{eff}})$. Here, we have made use of the following integral identity\footnote[3]{The authors have noted an error in the corresponding identity in the 7th edition of \textit{Gradshteyn \& Ryzhik--- Table of integrals, series, and products}, which has been duly corresponded for errata.} involving Whittaker functions\cite{buchholz2013confluent},
\begin{equation}
\begin{aligned}
\int_0^{\infty}dv~\exp\left[-\left(\frac{a+b}{2}\right)t \cosh v\right]I_{2i\nu}\left(t\sqrt{ab}\sinh v\right)
=\frac{\Gamma\left(\frac{1}{2}+i\nu\right)}{t\sqrt{ab}~\Gamma(1+2i\nu)}~W_{0,i\nu}\left(at\right)M_{0,i\nu}\left(bt\right)~,
\end{aligned}
\end{equation}
with the following identifications: $e^{y_0}=a$, $e^{y_1}=b$, and $t=(4/r_{+})$ and the fact that \cite{abramowitz1948handbook},
\begin{equation}
W_{0,i\nu}(z)=\sqrt{\frac{z}{\pi}}~K_{i\nu}\left(\frac{z}{2}\right)~~\text{and},~~M_{0,i\nu}(z)=2^{2i\nu +\frac{1}{2}}~\Gamma(1+i\nu)\sqrt{\frac{z}{2}}~I_{i\nu}\left(\frac{z}{2}\right)~.
\end{equation}
On substituting back the above result, we obtain the expression used in the main text.

\bibliographystyle{utphys1}
\bibliography{mybib}
\end{document}